\documentclass[showpacs,prb,amsmath,amssymb]{revtex4}
\usepackage{graphicx} 
\usepackage{color} 
\usepackage{bm}
\usepackage{psfrag}
\long\def\comment#1{}

\begin{document}

\title{Exact magnetization plateaus and phase transitions in 
  spin-$S$ Heisenberg antiferromagnets in arbitrary dimensions}

\author{V.~Ravi Chandra and Naveen Surendran}
  \affiliation{Centre for High Energy Physics, Indian Institute of
  Science, Bangalore - 560 012, India}

\begin{abstract}
We generalize a class of Heisenberg antiferromagnets in one, two and
three dimensions, which have been shown to exhibit magnetization
plateaus for spin-$\frac{1}{2}$. In a certain parameter range of the
general model, which is formally defined in $D$ dimensions, we obtain
the exact ground state(s) in the presence of an external magnetic
field for arbitrary values of spin. In this range, the magnetization
remains a constant as a function of the external field, except at some
special values of the field where there is a jump from one plateau to
the next. The plateaus are formed at certain specific fractions of the
full magnetization which are determined by the spin and the
lattice. Our general spin-$S$ result reproduces the known cases for
spin-$\frac{1}{2}$ in various lattices. Furthermore, we argue that
outside the exact regime, the mechanism for the plateau formation is
different. This results in first order phase transitions along some of
the plateaus as the coupling constant is varied. We rigorously show
the existence of such transitions for some particular cases. Finally,
we numerically analyze a spin-$1$ model in one dimension using exact
diagonalization to obtain its complete phase diagram. It agrees with
our analytic results.

\end{abstract}


\maketitle

\section{\label{Intro} Introduction}
\begin{figure}[h]
\psfrag{j}[l,b][l,b]{$J$}
\psfrag{jp}[l,b][l,b]{$J^\prime$}
\psfrag{1i}[c,b][c,b]{$(1,i)$}
\psfrag{2i}[c,b][c,b]{$(2,i)$}
\begin{center}
\includegraphics[width= .4\textwidth]{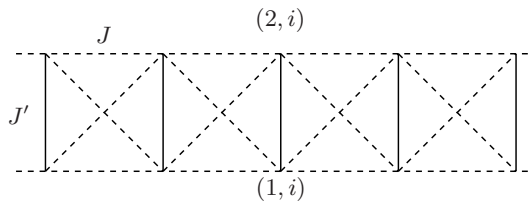}
\end{center}
\caption{\label{gelf-ladder}The Gelfand ladder. The full lines
represent the coupling $J^\prime$ and the dashed lines $J$.}
\end{figure}

Many low-dimensional spin systems, especially in one dimension,
respond discontinuously to the variations of an external magnetic
field \cite{CabraGrynbergHonecker}. In such systems, magnetization
curve forms plateaus as a function of the field. In most cases,
plateaus are continuously connected to one another, whereas a more
interesting scenario is when there are abrupt jumps between two
plateaus. This usually happens at high fields in systems with
localized magnon excitations\cite{SchulenburgHoneckerSchnack}. There
is a class of models for which magnetization plateaus of the second
type are formed via a simple mechanism. The basic feature of these
models is that they consist of units of spins coupled together such
that the total spin of each unit is conserved \cite{Gelfand,
BoseGayen, HoneckerMilaTroyer, MullerSinghKnetter, Sutherland}. The
first of its kind was introduced and studied at zero field by Gelfand
\cite{Gelfand}. The system consists of coupled dimers with the
following Hamiltonian (see Fig. \ref{gelf-ladder}),
\begin{equation}
\label{h-gelfand}
H_l = J^\prime \sum_i {\bf S}_{1,i} \cdot {\bf S}_{2,i}
+ J \sum_i \left({\bf S}_{1,i} + {\bf S}_{2,i} \right) \cdot 
\left({\bf S}_{1,i+1} + {\bf S}_{2,i+1} \right).
\end{equation}

Note that the total spin on each dimer $(S_{1,i}^a + S_{2,i}^a)$ is
conserved. A whole lot of eigenstates can be trivially constructed
for the above Hamiltonian. Any state in which each dimer has a
specific total spin and is maximally polarized in the $z$-direction
will be an eigenstate. Then, one expects the ground state to be a
product of dimer-singlets for $J^\prime >> J$. A lower bound on
$J^\prime$ such that the dimer-singlet is the ground state has been
obtained by rewriting the Hamiltonian as sum of units consisting of
three spins \cite{ShastrySutherland}, and is given by,
\begin{equation}
\label{lower-bound}
J^\prime  \ge \left\{ \begin{array}{ll}
2J, & ~~~~~~\mbox{for}~ S=\frac{1}{2}, \\
2J(S+1), &~~~~~~ \mbox{for}~ S \ge 1.
\end{array} \right.
\end{equation}
For $S=\frac{1}{2}$, the ground state in fact remains to be the
dimer-singlet for $J^\prime > 1.4 J $. For $J^\prime < 1.4 J$, all the
dimers go into triplets and form the ground state of a spin-$1$
uniform chain.

The lower-bound for $J^\prime$ given in \eqref{lower-bound} suggests
that, the higher the spin, the stronger the dimer bond needs to be for
the dimer-singlet to be the ground state. But it turns out that the
dimer-singlet is quite robust even for very large spins. In this
paper, we show that the lower bound on $J^\prime $ can be pushed down
to $2J$ for arbitrary values of spin, {\it i.e.}, the bound given by
(\ref{lower-bound}) for $S=\frac{1}{2}$ is in fact true for all values
of $S$.

Honecker, Mila and Troyer\cite{HoneckerMilaTroyer} (from hereon
referred to as HMT) have numerically studied the $S= \frac{1}{2}$
system in the presence of magnetic field. For $J^\prime > 2J$, the
system starts with the dimer-singlet ground state at zero field and
the magnetization per site ($M$) is zero. The system remains in the
dimer-singlet phase till a transition field $B_{c_1} = J^\prime$, at
which one set of alternate dimers form triplets polarized in the
direction of the magnetic field ($M = \frac{1}{2}$). For further
increase in the field strength, this state remains to be the ground
state till a second transition field $B_{c_2} = 2J + J^\prime$, at
which the rest of the dimers also become triplets and the system is
fully polarized ($M=1$). This behaviour has also been inferred on the
basis of a strong coupling analysis \cite{Mila}.  For $J^\prime < 2J$,
gapless phases come into play and the magnetization curve starts
developing tails between plateaus before the latter disappear
altogether. HMT also studied a three-leg ladder and obtained abrupt
jumps between plateaus in some range of the parameter. Similar results
for $S=\frac{1}{2}$ have been obtained for a modified
Shastry-Sutherland model in two dimensions\cite{MullerSinghKnetter}
and a three-dimensional version of Gelfand ladder\cite{Sutherland}.

In this paper, we define a general model in arbitrary dimensions which
includes all the models mentioned above. In a certain regime of the
parameter space, we obtain the exact ground state(s) in presence of
magnetic field for arbitrary values of spin. The magnetization curve
forms plateaus and they jump abruptly from one to another at
certain special values of the field. The magnetization on the plateaus
are some rational fractions of the full magnetization, which depend on
the underlying spin and the lattice.

Outside the exact regime, we argue that the plateaus will survive in
the immediate neighbourhood, quite possibly developing tails between
successive ones. But the mechanism for plateau formation in this case
is different from the exact regime and this results in first order
phase transitions along some of the plateaus. We rigorously show the
existence of such transitions for some special cases. Finally, we
numerically analyse the ladder defined in \eqref{h-gelfand} for $S=1$
using exact diagonalization. We obtain the complete phase diagram
which confirms our analytic results.


\section{\label{klarge} The General model and the exact ground-states}

HMT have defined a general $n$-leg ladder in one dimension. We
generalize further and formally define Hamiltonians on an arbitrary
bipartite lattice in $D$ dimensions, where a bunch of $n$ spins live
on each lattice-site and interact among each other as well as with the
spins living on neighbouring sites. The reason for writing down the
most general Hamiltonian is that, most of our analysis is independent
of details like dimension, lattice and spin.

Let ${\bf S}_{\mu,{\bf x}}$ (\mbox{$\mu = 1,2,\cdots,n$}) be the spins
  living at the site ${\bf x}$. Define,
\begin{equation}
\label{tvar}
{\bf T}_{\bf x} = \sum_{\mu =1}^{n} {\bf S}_{\mu, {\bf x}}.
\end{equation}
Then the Hamiltonian is defined as,
\begin{equation}
\label{gen-ham}
H = \frac{J^\prime}{2} \sum_{{\bf x}} {\bf T}_{{\bf x}}^2 + 
 J \sum_{\langle {\bf x}, {\bf y} \rangle} {\bf T}_{\bf x} \cdot
{\bf T}_{\bf y} - B \sum_{\bf x} T_{\bf x}^z,
\end{equation}
where $\langle ~,~ \rangle$ denote nearest neighbours. We take both
the couplings to be positive. The models mentioned previously are
particular cases of the above Hamiltonian. For the Gelfand ladder and
the three-leg ladder in HMT, the underlying lattice is a simple chain
and $n=2$ and $3$ respectively. Both for the modified
Shastry-Sutherland model\cite{MullerSinghKnetter} and the 3-$d$ model
of Sutherland\cite{Sutherland} $n=2$. For the former the underlying
lattice is square and for the latter it is one-fourth depleted cubic.

The variables $T_{\bf x}^a$'s are also $SU(2)$ generators, and they
can take the representations having the Casimir \mbox{${\bf T}_{\bf
x}^2 = j(j+1)$}, such that \mbox{$j = 0,1,\cdots,nS$}, when $nS$ is an
integer; and \mbox{$j = \frac{1}{2}, \frac{3}{2}, \cdots, nS$}, when
$nS$ is a half-odd integer. Thus the bunch of spins at each site
kinematically act respectively like a bosonic or fermionic rotor,
except that the values $j$ can take are truncated at $nS$. The
Hamiltonian in Eq. \eqref{gen-ham}, feels the value of the underlying
spin only through this truncation. This suggests that one can treat
the problem for general values of spin.

\begin{figure}[t]
\psfrag{j}[c,b][c,b]{$J$}
\psfrag{jp/2}[c,c][c,c]{$\frac{J^\prime}{2}$}
\begin{center}
\includegraphics[width= .5\textwidth]{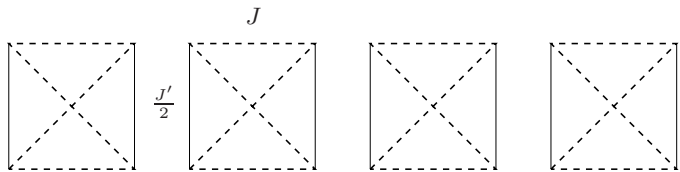}
\end{center}
\caption{\label{divide}Writing the ladder Hamiltonian as a sum over
units of two dimers.}
\end{figure}

Now we will use a method of {\it divide-and-conquer} to solve for the
ground state. If one can write $H$ as a sum over smaller units
$h_i$, such that there exists a state $|\psi_0 \rangle$ which is
simultaneously a ground state of each and every $h_i$, then $| \psi_0
\rangle$ will be a ground state of $H$. To this end, we rewrite
$H$ as a sum over all the bonds of the underlying lattice. Let
$i$ label the bonds and let $z$ be the co-ordination number of the
lattice. Then,
\begin{equation}
\label{divide-ham}
H = \sum_i h_i ,~~~~~~~~
h_i = \frac{J^\prime}{2z} \left({\bf T}_{1,i}^2 + 
{\bf T}_{2,i}^2 \right) + J {\bf T}_{1,i} \cdot {\bf T}_{2,i}
- \frac{B}{z} \left( T_{1,i}^z + T_{2,i}^z \right).
\end{equation}
where ${\bf T}_{1,i}$ and ${\bf T}_{2,i}$ are the two degrees of
freedom living on the bond $i$. Note that the same spin will be
shared by many bonds and hence labeled in more ways than one. For the
Gelfand ladder, the above break-up is schematically represented in Fig
\ref{divide}.

Now we will concentrate on a particular $h_i$, and find its ground
state and energy as a function of $B$. For notational convenience we
shall drop the index $i$ from hereon. After a rearrangement of terms, the
bond  Hamiltonian can then be rewritten as,
\begin{equation}
\label{h-divide2}
h =  \frac{1}{2z} \left(J^\prime - zJ \right) 
 \left({\bf T}_1^2 + {\bf T}_{2}^2 \right) + \frac{J}{2} 
\left( {\bf T}_1 + {\bf T}_{2} \right)^2
- \frac{B}{z} \left( T_1^z + T_{2}^z \right).
\end{equation}
The four terms in $h$ mutually commute and therefore the spectrum is
trivially solved. Let \mbox{$j_1(j_1+1)$}, \mbox{$j_2(j_2+1)$},
\mbox{$j(j+1)$} and $m$ be the eigenvalues of ${\bf T}_1^2$, ${\bf
T}_2^2$, \mbox{$({\bf T}_1+{\bf T}_2)^2$} and $( T_1^z + T_{2}^z)$
respectively. Here 
\mbox{$|(j_1 - j_2)| \le j \le (j_1 + j_2)$} and $-j \le m \le
j$. Then the spectrum for $h$ is given by,
\begin{equation}
\label{spectrum-divide}
E(j_1,j_2,j,m) = \frac{1}{2z} \left(J^\prime-zJ \right) 
 \Big(j_1(j_1+1) + j_2(j_2+1) \Big) + \frac{J}{2} ~j (j+1)
- \frac{B}{z} ~m.
\end{equation}
Since the magnetic field couples to a conserved quantity, the problem
of finding the ground state at a given field strength reduces to
finding the lowest energy state in each $j$-sector at zero field, {\it
i.e.}, we need to minimize the field-independent part of $E$ with
respect to $j_1$ and $j_2$. For non-zero field, the degeneracy arising
from the $SU(2)$ symmetry will be lifted and the ground state will
have $m=j$. Next we consider the case $J^\prime > zJ$, for which the
coefficient of the first term in $E$ is positive. 

When $nS$ is an integer, \mbox{$j_1,j_2 = 0,1, \cdots ,nS$} and for
half-odd integer values, \mbox{$j_1,j_2 = \frac{1}{2}, \frac{3}{2},
\cdots, nS$}. In both the cases, \mbox{$j = 0,1, \cdots, 2nS$}.
Anticipating the solutions, we divide all possible cases into two:
\begin{enumerate}

\item  Even $j$, integer $nS$;\\ Odd $j$, half-odd integer $nS$.

\item  Odd $j$, integer $nS$;\\ Even $j$, half-odd integer $nS$.

\end{enumerate}
For Case $1$, $E$ is minimized by the choice,
\begin{equation}
\label{jeven}
j_1= j_2= \frac{j}{2}.
\end{equation}
And for Case $2$, the minimum of $E$ is when,
\begin{equation}
\label{jodd}
j_1= \frac{(j-1)}{2},~~ j_2= \frac{(j+1)}{2}.
\end{equation}
There is one exception to the above general rule, though. When $nS$ is
half-odd integer and $j=0$, Eqs. \eqref{jodd} give $j_1 =-\frac{1}{2}$
and $j_2= \frac{1}{2}$. This is not an admissible solution and the
correct solution is, \mbox{$j_1 = j_2 = \frac{1}{2}$}. We will have
more to say about this exceptional case later.

The ground state energies in the presence of field corresponding to
the solutions \eqref{jeven} and \eqref{jodd} are respectively given by,
\begin{eqnarray}
\label{e0-even}
E_0^{(1)} &=& \frac{1}{4z} \left(J^\prime-zJ \right) 
 j(j+2) + \frac{J}{2} ~j (j+1) - \frac{B}{z} ~j, \\
\label{e0-odd}
E_0^{(2)} &=& \frac{1}{4z} \left(J^\prime-zJ \right) 
 (j+1)^2 + \frac{J}{2} ~j (j+1) - \frac{B}{z} ~j.
\end{eqnarray}
Let \mbox{$|l,m \rangle_\alpha$} denote the state with eigenvalues
$l(l+1)$ and $m$ for the operators ${\bf T}_\alpha^2$ and $T_\alpha^z$
respectively. Then the corresponding ground states are,
\begin{eqnarray}
\label{gs-even}
|\psi^{(1)}_0(j) \rangle &=& |j/2,~j/2 \rangle_1 \otimes
|j/2,~j/2 \rangle_2, \\
\label{gs-odd}
|\psi^{(2)}_0(j) \rangle &=& |(j-1)/2,~(j-1)/2 \rangle_1 \otimes
|(j+1)/2,~(j+1)/2 \rangle_2.
\end{eqnarray}
Since the field-independent part of $E_0^{(1,2)}$ are monotonically
increasing functions of $j$, as $B$ is increased, $j$ will go up
sequentially in steps. The value of the field at which the total spin
goes up to \mbox{$(j+1)$} from $j$ can be obtained from
Eqs. \eqref{e0-even} and \eqref{e0-odd}.  For Case $1$, the transition
field is,
\begin{equation}
\label{bcrit-even}
B_{c_j}^{(1)} = \frac{(J^\prime + z J)}{2}~ j + J^\prime.
\end{equation}
For Case $2$, it is given by,
\begin{equation}
\label{bcrit-odd}
B_{c_j}^{(2)} = \frac{(J^\prime + z J)}{2}~ (j+1).
\end{equation}
This completes the solution of the bond-Hamiltonian ($h$) for
$J^\prime > zJ$. The important point to note is that the ground states
\eqref{gs-even} and \eqref{gs-odd} are product states of the two sites
connected by the bond. In the context of the full lattice, this means
that we can put together these states living on the bonds and
construct a state which will be a simultaneous ground state of $h_i$
for every $i$. Consequently, it will be a ground state of the full
Hamiltonian, $H$.

For Case $1$, the ground state consists of all dimers having total
spin $j/2$ and maximal polarization in the $z$-direction, {\it i.e.},
\mbox{${\bf T}_\alpha^2 = (j/2)(j/2+1)$} and \mbox{$T_\alpha^z =
j/2$}. (Note that $j$ denotes the total spin on the bonds, which are
not good quantum numbers. Here we merely use $j$ to label the ground
states). Since the ground state is unique for individual bond
Hamiltonians, this state has to be the unique ground state of the full
Hamiltonian. The state can be written as,
\begin{equation}
\label{gs-ladder-even}
|\Psi^{(1)}_0(j) \rangle = \prod_{{\bf x}} |j/2,j/2 \rangle_{\bf x}
\end{equation}
For Case $2$, the ground state can be written as,
\begin{equation}
\label{gs-ladder-odd}
|\Psi^{2}_0(j) \rangle = \prod_{{\bf x} \in A} |(j-1)/2, (j-1)/2
\rangle_{\bf x} 
\prod_{{\bf y} \in B} |(j+1)/2, (j+1)/2 \rangle_{\bf y},
\end{equation}
where $A$ and $B$ denote the two sub-lattices. The above ground state
breaks the translation symmetry and is therefore doubly degenerate,
the other ground state being the one in which the two sub-lattices are
interchanged. 
Eqs. \eqref{gs-ladder-even} and \eqref{gs-ladder-odd} give the ground
states for all values of $j$ ranging from $0$ to $2nS$.  The
fractions of the total magnetization at which the plateaus form are,
\begin{equation}
\label{fractions}
M=\frac{j}{2nS},
\end{equation}
and the transition fields are given by Eqs. \eqref{bcrit-even} and
\eqref{bcrit-odd}. At each jump of the magnetization, dimers belonging
to one of the sub-lattices alternately get excited to a total spin one
unit higher. These plateau transitions are similar to the macroscopic
magnetization jumps discussed by Schulenburg {\it et. al.}
\cite{SchulenburgHoneckerSchnack}. The difference being that, in our
case the localization of magnons arises from conservation laws.

The above solution for the ground state hinges on one crucial aspect -
the ground state of the bond Hamiltonian ($h$) is a product state on
the two sites. For half-odd integer $nS$ and $j=0$, we have seen that
the ground state of $h$ has $j_1 = j_2 = \frac{1}{2}$. This is not a
product state and thus cannot necessarily be satisfied for all
bonds. When $nS$ is half-odd integer, the minimum spin on any site is
$\frac{1}{2}$ and therefore at zero field the system will be in the
ground state of spin-$\frac{1}{2}$ system on the underlying
lattice. This is gapless, even in one-dimension\cite{Bethe}, and
therefore the magnetization will vary continuously with the field. For
$j=1$, the solution again has $j_1 = j_2 = \frac{1}{2}$. This is the
fully polarized state of the spin-$\frac{1}{2}$ system. The field at
which it becomes fully polarized can be determined by the gap of
one-magnon states ({\it i.e.}, all spins except one pointing up). For
a hyper-cubic lattice in $D$ dimensions, this is calculated to be,
\begin{equation}
\label{1/2gapless}
\frac{B_c}{J} = 4SD
\end{equation}

Our general expressions for the magnetization fractions
(Eq. \eqref{fractions}) and transition fields
(Eqs. \eqref{bcrit-even}, \eqref{bcrit-odd} and \eqref{1/2gapless})
check correctly with the results in HMT for spin-$\frac{1}{2}$, two and
three-leg ladders.

A comment is in order here regarding the ground states at plateau
transitions. Suppose the field is tuned to $B_{c_j}$, {\it i.e.} the
transition field at which $j$ goes up to $j+1$. Also let $j$ be even
(a similar argument holds for odd $j$ as well). Then the state of each
bond can be any one of the following three
states.
\begin{eqnarray}
|\tilde{\psi}_0(j) \rangle_1 &=& |j/2,~j/2 \rangle \otimes 
|j/2,~j/2 \rangle, \\
|\tilde{\psi}_0(j) \rangle_2 &=& |j/2,~j/2 \rangle \otimes 
|(j/2 + 1),~(j/2+1) \rangle, \\
|\tilde{\psi}_0(j) \rangle_3 &=& |\left(j/2+1 \right),~
\left(j/2+1 \right) \rangle \otimes 
|j/2,~j/2 \rangle.
\end{eqnarray}
This gives rise to a degree of degeneracy which grows exponentially with
the system size. In HMT, it is mentioned that the ground state of the
spin-$\frac{1}{2}$ ladder is either all triplets (not necessarily polarized)
or a mix of singlets and fully polarized triplets. They make an
exception for those special values of the field at which there is a
jump in magnetization. The above analysis shows that even at
transition fields the ground states belong to the afore-mentioned set.

\section{\label{beyond}Beyond the exact regime}

The {\it divide-and-conquer} method we employed in the previous
section to find the exact ground state worked because the ground state
of the bond Hamiltonian ($h$) is a product
state. This is no longer true for $J^\prime < zJ$. Then the ground
state of $h$ for a particular $j$ will have,
\begin{equation}
\label{3-dimer-gs}
{\bf T}_1^2= nS(nS+1),~~{\bf T}_2^2 = nS(nS+1),~~
({\bf T}_1 + {\bf T}_2)^2 = j(j+1),~~(T_1^z + T_2^z)= j,
\end{equation}
This state is not a product state unless $j=nS$. But this does not
necessarily mean that the product states cease to be the ground state
of $H$ for all $J^\prime < zJ$. Now we will argue that the product
states, and hence the magnetization plateaus, do indeed survive for
a while as $J^\prime$ goes below $zJ$. Significantly, the
solutions are not same as that for $J^\prime > zJ$ as given by
Eqs. \eqref{gs-ladder-even} and \eqref{gs-ladder-odd}. The new
solutions also form plateaus at the same fractions and it results in a
first order phase transition on each plateau (with certain exceptions)
at $J^\prime = zJ$.

For the proof of the existence of product state solutions for
$J^\prime<zJ$, we do the following. First we will show that, at
$J^\prime =zJ$, the only solutions are a finite number of product
states, except at the transition fields. Then, by assuming these states to
be gapped, it will follow that there exists a positive non-zero
$\Delta$ such that the ground state at \mbox{$(z - \Delta)J < J^\prime
< zJ$} will be among the ground states at \mbox{$J^\prime = zJ$}.

At $J^\prime =zJ$, the bond Hamiltonian in Eq. \eqref{h-divide2}
becomes,
\begin{equation}
\label{h-dimer2}
h =  \frac{J}{2} \left( {\bf T}_1 + {\bf T}_{2} \right)^2
- \frac{B}{z} \left( T_1^z + T_{2}^z \right).
\end{equation}
The above Hamiltonian sees only the total spin of the bond, but
is independent of the spins of the sites. Then the ground
states of a given $j$-sector are highly degenerate - there is a
freedom to choose $j_1$ and $j_2$ as long as they combine to form
total spin $j$. Among these states, there is a subset given by,
\begin{equation}
\label{gs@k=2}
|\psi_0 \rangle_p = |l,l \rangle_1 \otimes |(j-l),(j-l) \rangle_2,
~~~~l=0,1,\cdots,j ,
\end{equation}
which constitutes all the only product states. Next
we will show that the ground states of the full system at
$J^\prime =zJ$ can be written as product states on the sites. Let
us consider two neighbouring bonds - $a$ connecting sites $1$ and $2$,
and $b$ connecting sites $2$ and $3$. Also let $h_a$ and $h_b$ be the
corresponding bond Hamiltonians. Then, at $J^\prime
=zJ$ any ground state, of \mbox{$(h_a + h_b)$} has
to satisfy the following conditions.
\begin{eqnarray}
\label{constraint1}
\left( {\bf T}_1 + {\bf T}_{2} \right)^2 |\phi_0 \rangle
&=& j(j+1) |\phi_0 \rangle , \\
\label{constraint2}
\left( {\bf T}_1^z + {\bf T}_{2}^z \right) |\phi_0 \rangle
&=& j|\phi_0 \rangle ,\\
\label{constraint3}
\left( {\bf T}_2 + {\bf T}_{3} \right)^2 |\phi_0 \rangle
&=& j(j+1) |\phi_0 \rangle , \\
\label{constraint4}
\left( {\bf T}_2^z + {\bf T}_{3}^z \right) |\phi_0 \rangle
&=& j|\phi_0 \rangle ,
\end{eqnarray}
where $j$ is determined by the field, $B$. In the sub-space of fixed
eigenvalues for ${\bf T}_1^2$, ${\bf T}_2^2$ and ${\bf T}_3^2$,
Eqs. \eqref{constraint1} and \eqref{constraint2} imply
that,
\begin{equation}
|\phi_0 \rangle = |\phi_0 \rangle_{12} \otimes |\phi_0 \rangle_3,
\label{phi12}
\end{equation}
where, $|\phi_0 \rangle_{12}$ lives in the combined Hilbert space of
the sites $1$ and $2$, and $ |\phi_0 \rangle_3$ lives in the Hilbert
space of the site $3$. This follows from the fact that for fixed
values of $j_1$ and $j_2$, there is a unique state in the Hilbert
space of sites $1$ and $2$ which satisfies Eqs. \eqref{constraint1} 
and \eqref{constraint2}. Similarly Eqs. \eqref{constraint3} and 
\eqref{constraint4} imply that,
\begin{equation}
|\phi_0 \rangle = |\phi_0 \rangle_{1} \otimes |\phi_0 \rangle_{23},
\label{phi23}
\end{equation}
where, $|\phi_0 \rangle_1$ lives in the Hilbert space of the site $1$
and $|\phi_0 \rangle_{23}$ lives in the combined Hilbert space of the
sites $2$ and $3$. Finally, Eqs. \eqref{phi12} and \eqref{phi23}
imply that,
\begin{equation}
|\phi_0 \rangle = |\phi_0 \rangle_{1} \otimes |\phi_0 \rangle_{2}
\otimes |\phi_0 \rangle_{3},
\label{phi123}
\end{equation}
{\it i.e.}, $|\phi_0 \rangle$ is a product state of the three sites.
Since every site is part of at least two bonds, this necessarily
implies that the ground states of $H$ at $J^\prime = zJ$ are product
states. These ground states can be written as,
\begin{equation}
\label{gs-ladder-k2}
|\Psi_0(J^\prime=zJ,j) \rangle = \prod_{{\bf x}\in A} |l,l
\rangle_{\bf x} \otimes \prod_{{\bf y}\in B} |j-l,j-l \rangle_{\bf y},
~~~~l=0,1,\cdots,j ,
\end{equation}
What we have shown now is that, though the bond Hamiltonian ($h$)
admits solutions which are not product states at $J^\prime = zJ$, such
solutions are not admissible for the full Hamiltonian ($H$).

At transition fields, there is a greater degree of degeneracy. Then
the constraint on the ground state is more relaxed - the total spin on
a bond can be $j$ or $j+1$. This gives rise to a degeneracy which is
local and hence grows exponentially with the system size. Away from
the transitions fields, the ground states are those given by
Eq. \eqref{gs-ladder-k2} and are not related to each other by any
local transformation. It is then fair to assume that there is a gap to
excitations.

Now let us see why gap at $J^\prime=J$ implies the existence of
product ground state for $(z-\Delta)J<J^\prime<2J$, for some positive
$\Delta$.  The three terms in $H$ in Eq. \eqref{gen-ham}
mutually commute. This means that one can choose the eigenstates to be
independent of the three parameters $J^\prime$, $J$ and $B$. The
energy spectrum will depend linearly on all the three parameters and
all transitions of the ground state are via level crossings. (The
transitions will be first order unless there is a continuum of level
crossings - a possibility in the thermodynamic limit). A gap to the
ground states at $J^\prime=zJ$ would then imply that there exists some
non-zero positive number $\Delta$, such that for \mbox{$(z-\Delta)J <
J^\prime < zJ$} the ground state(s) will be among those at
$J^\prime=zJ$. Among the states in Eq. \eqref{gs-ladder-k2}, the one
with the lowest energy for $J^\prime < zJ$ is given by,
\begin{eqnarray}
\label{gs<1}
|\Psi^<_0(j) \rangle &=& \prod_{i \in A} |j,j \rangle \otimes
\prod_{k \in B} |0,0 \rangle,~~~~\mbox{for $j \le 2S$,} \\
\label{gs<2}
|\Psi^<_0(j) \rangle &=& \prod_{i \in A} |2S,2S \rangle \otimes
\prod_{k \in B} |(j-2S),(j-2S) \rangle,~~~~\mbox{for $j>2S$}.
\end{eqnarray}
These states also have the same fractions for magnetization as
before. But, now the system prefers to put all the spin into sites in
one sub-lattice, if that is possible. This is in contrast to the
situation for $J^\prime > zJ$ where it is energetically favourable to
distribute the spin equally among the two sub-lattices. At $j=2S$, one
sub-lattice will be saturated and then the other sub-lattice start
getting excited to higher and higher spins.

In the above states, when $j>2S$, sites in one sub-lattice have spin
$2S$ and those in the other sub-lattice have spin $(j-2S)$. On the
plateau, each spin is maximally polarized in the $z$-direction. Below
a critical field, gapless phase of the same set of spins will have
lower energy. The critical field, in units of $J$, will be the gap in
the spectrum of one-magnons (states with all spins but one maximally
polarized, the odd one having $z$-component one less than
maximum). For a hyper-cubic lattice in $D$-dimensions consisting of
spins $j_1$ in one sub-lattice and $j_2$ in the other, the critical
field is given by,
\begin{equation}
\label{bcrit-gapless}
\frac{B_{crit}}{J} = 2D(j_1+j_2).
\end{equation}
HMT have numerically found such lines of transition for
$S=\frac{1}{2}$, two and three legged ladders and the formula above is
consistent with their results.  For the two-leg ladder, the transition
occurs when the fully polarized effective spin-$1$ chain becomes
gapless.  Then \mbox{ $j_1=j_2=1$ }, and \mbox{$B_{crit}=4J$}. For
the three-leg ladder there are three cases - $a$)
\mbox{$j_1=\frac{1}{2},$} \mbox{$j_2=\frac{1}{2}$}, $b$)
\mbox{$j_1=\frac{1}{2},$} \mbox{$j_2=\frac{3}{2}$} and $c$)
\mbox{$j_1=\frac{3}{2}$}, \mbox{$j_2=\frac{3}{2}$}. Then the
transitions occur respectively at \mbox{$B_{crit} = 2J,~4J~\mbox{and}~
6J$}.

To summarize our results thus far, we first solved for the ground
states of the general spin-$S$ model in arbitrary dimensions for
$J^\prime > zJ$. The product state solutions we obtain exhibit
magnetization plateaus. Then we argued that there exist product state
solutions and hence magnetization plateaus even as $J^\prime$ goes
below $zJ$, but the mechanism for plateau formation is different from
the case $J^\prime >zJ$. Thus, on a plateau, there are {\it first order
phase transitions at $J^\prime=zJ$}. 

HMT have obtained a similar first order phase transition on the
plateau for the three-leg, $S=\frac{1}{2}$ model. In their case the
transition occurs on $M=\frac{1}{3}$ plateau from a fully polarized
$S=\frac{1}{2}$ chain to an alternating chain consisting of spins
$\frac{1}{2}$ and $\frac{3}{2}$. The latter is not a simple product
state.

The plateau solutions for $J^\prime >2J$, given by
Eqs. \eqref{gs-ladder-even} and \eqref{gs-ladder-odd}, and that for
$J^\prime < 2J $, given by Eqs. \eqref{gs<1} and \eqref{gs<2},
coincide for $j=0,~1,~(2nS-1)~\mbox{and}~2nS.$ For $S=\frac{1}{2}$
ladder ($n=2$), all the plateaus belong to the above set and therefore
there is no first order transition at $J^\prime=2J$ along any of the
plateaus.

From now we will concentrate on the two-leg ladder. First we will show
for some higher spin cases that first order transitions along the
plateaus at $J^\prime = 2J$ does indeed exist.

\section{Phase transitions for the ladder at $J^\prime = 2J$}

\begin{figure}
\psfrag{2i}[c,b][c,b]{$2i$}
\psfrag{2i-1}[c,b][c,b]{$(2i-1)$}
\psfrag{2i+1}[c,b][c,b]{$(2i+1)$}
\begin{center}
\includegraphics[width = .6 \textwidth]{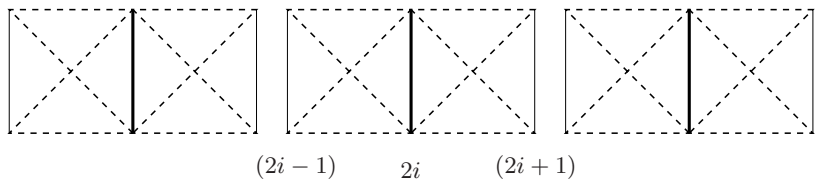}
\end{center}
\caption{\label{3-dimer}Writing the Hamiltonian as a sum over units of three
  dimers. Bold full lines represent bonds of strength $J^\prime$,
  thin full lines $J^\prime / 2$ and dashed lines $J$.}
\end{figure}

We have seen that the ground states of the bond Hamiltonian are not
product states for $J^\prime < 2J$.  Therefore, to look for product
state solutions in this range, we try a different {\it
divide-and-conquer} scheme involving bigger units. To write $H_{l}$
as sum over units containing three dimers then suggests itself ( see
\mbox{Fig. \ref{3-dimer}}),
\begin{eqnarray}
\label{h-ladder-3dimer}
H_{l} &=& \sum_i \tilde{h}_i, \\
\label{h-3dimer}
\tilde{h}_i &=& \frac{J^\prime}{4} \left({\bf T}_{2i-1}^2 +
2 {\bf T}_{2i}^2 +  {\bf T}_{2i+1}^2 \right) 
+ J {\bf T}_{2i} \cdot ( {\bf T}_{2i-1} + {\bf T}_{2i+1})
- \frac{B}{2}  \left( T_{2i-1}^z + 2~T_{2i}^z + T_{2i+1}^z \right).
\end{eqnarray}
As before, we drop the index $i$ and write the $3$-dimer Hamiltonian
as,
\begin{equation}
\label{h-3dimer1}
\tilde{h} = \frac{J^\prime}{4} \left({\bf T}_{1}^2 +
2 {\bf T}_{2}^2 +  {\bf T}_{3}^2 \right) 
+ J {\bf T}_{2} \cdot ( {\bf T}_{1} + {\bf T}_{3})
- \frac{B}{2}  \left( T_{1}^z + 2~T_{2}^z + T_{3}^z \right).
\end{equation}
${\bf T}_{1}^2$, ${\bf T}_{2}^2$, ${\bf T}_{3}^2$, \mbox{$({\bf T}_{1}
+ {\bf T}_{3})^2$} and \mbox{$({\bf T}_{1}^z + {\bf T}_{2}^z + {\bf
T}_{3}^z )$} are mutually commuting conserved quantities and this
essentially reduces the problem to that of two (variable) interacting
spins coupled to a magnetic field of unequal strengths at the two
sites. Unfortunately, the spectrum cannot be found analytically for
arbitrary values of spin.  

We have numerically diagonalized $\tilde{h}$ for spins up to $3$.
We have found that the product state solutions given in
Eqs. \eqref{gs<1} and \eqref{gs<2} do indeed exist for some $J^\prime <
2J$. (This diagonalization can be easily performed for higher spins
also, but we chose to restrict it to reasonable values).
Note that the existence of the product state solutions for $\tilde{h}$
necessarily implies their existence for the full ladder. This in turn
confirms our prediction of first order transitions on a plateau at
$J^\prime = 2J$.

Next we study the particular case $S=1$. We numerically
analyse it to obtain the phase diagram in the complete parameter space
and thus confirm our analytic results.

\section{Spin-$1$ ladder}

\begin{figure}
\begin{center}
\includegraphics[width = .55 \textwidth]{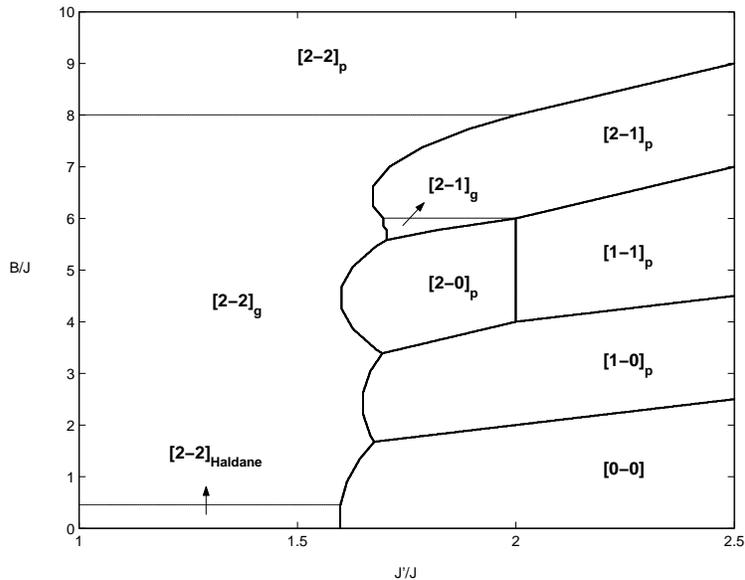}
\end{center}
\caption{\label{phasediagram} Phase diagram for the spin-$1$
  ladder. Thick and thin lines denote first and second
  order transitions respectively.}
\end{figure}
\begin{figure}
\begin{center}
\includegraphics[width = .5 \textwidth]{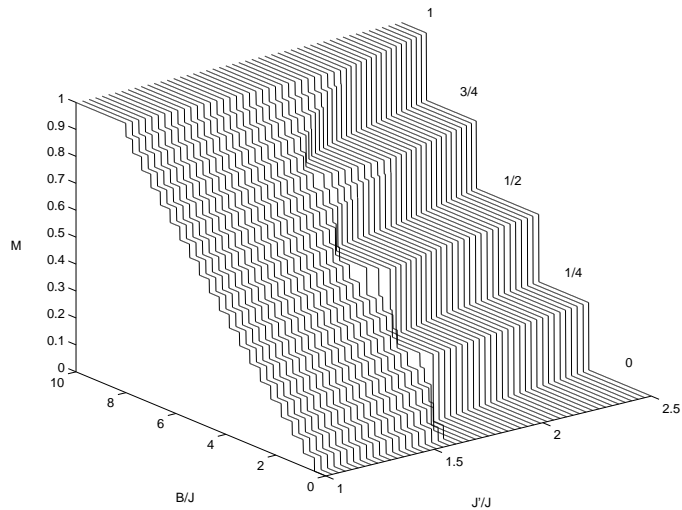}
\end{center}
\caption{\label{magcurves} Magnetization curves for the spin-$1$
  ladder. The fractional values of magnetization are denoted on the
  plateaus.}
\end{figure}
The spin-$\frac{1}{2}$ ladder was numerically studied in great detail by HMT
to obtain the complete phase diagram. Our results for $J^\prime >2J$ and
the arguments for the existence of the plateaus beyond the exact
regime are consistent with their phase diagram. As we mentioned in the
previous section, to obtain the new phase predicted by us, one needs
to look at spins higher than $\frac{1}{2}$. In this section we present the
results of an exact diagonalization analysis of the spin-$1$ ladder.

Since the individual dimer spins are conserved, we need to analyse
only effective spin chains consisting of spins of varying values at
each site. We have employed the Lanczos algorithm to exactly
diagonalize the Hamiltonian and obtain the ground state energies. We
have worked with $10$ dimers ($20$ spins) and used periodic boundary
conditions. For the finite system that we studied, we have checked
that all the relevant states are such that the spins of the dimers
belonging to a given sub-lattice are the same. There are exceptions to
this at plateau transitions, arising from the local degeneracy which
has been discussed in Section \ref{klarge}. In the phase diagram (see
Fig. \ref{phasediagram}), $[j_1-j_2]_p$ and $[j_1-j_2]_{g}$
respectively denote the fully polarized and the gapless phases of an
alternating chain consisting of spins $j_1$ and $j_2$. In
Fig. \ref{magcurves}, the magnetization is plotted as a function of
$B/J$ for various values of $J^\prime / J$.

For $J^\prime > 2J$ the analytic results from \mbox{Section
\ref{klarge}} are very accurately reproduced. Since all the relevant
states are product states, there are no finite-size corrections and
agreement with the analytic results is exact up to numerical
inaccuracies. There are five plateaus at the fractions
$0,~\frac{1}{4},~\frac{1}{2},~\frac{3}{4}$ and $1$. The successive
transitions occur at the following values of the field,
\begin{eqnarray}
&&\frac{B_{c_1}}{J} = \frac{J^\prime}{J},~~~~~~~~~~~~~~~~~
\frac{B_{c_2}}{J} = 2 + \frac{J^\prime}{J},\\
&&\frac{B_{c_3}}{J} = 2 + 2~\frac{J^\prime}{J},~~~~~~~
\frac{B_{c_4}}{J} = 4 + 2~\frac{J^\prime}{J}.
\label{plateau}
\end{eqnarray}
On the $M=\frac{1}{2}$ plateau, $[1-1]_p$ phase gives way to $[2-0]_p$
 at $J^\prime = 2J$, as we expect from our analysis in \mbox{Section
 \ref{beyond}}. From $[2-1]_p$ and $[2-2]_p$ phases there are second
 order transitions to the corresponding gapless phases $[2-1]_{g}$
 and $[2-2]_{g}$ respectively. The lines of these transitions,
 obtained from Eq. \eqref{bcrit-gapless}, are given by
 $B_{crit}=6J~\mbox{and}~8J$ respectively. These lines meet the lines
 separating plateaus, given in Eq. \eqref{plateau} at $J^\prime=2J$.

For $J^\prime \ll 2J$, all the dimers have spin $2$. Then the ground
state at zero field is the Haldane phase of the uniform spin-$2$
chain\cite{Haldane, ParkinsonBonner}. When the field is turned on, the
magnetization remains zero till a critical field at which the system
becomes gapless\cite{LouQinSu}. Then with further increase in field, the
magnetization continuously increases and finally reaches saturation
($[2-2]_p$). In the plot for magnetization curves in Fig
\ref{magcurves}, there are discernible steps even in the gapless
phases $[2-1]_g$ and $[2-2]_g$. These are spurious jumps arising from
finite-size effects. To obtain more accurate results, it will be
necessary to employ techniques like DMRG.

In the intermediate range, {\it i.e.} around $J^\prime \approx 1.6 J$,
just like in the $S=\frac{1}{2}$ case in HMT, with changing field, the
system goes in and comes out of various gapped and gapless phases. At
zero field, $[0-0]$ phase goes into spin-$2$ Haldane phase at
$J^\prime \approx 1.5975 J$.

For higher spins, there will be $(4S+1)$ plateaus for $J^\prime > 2J$
as discussed in Section \ref{klarge}. First order transitions of the
type, $[1-1]_p$ to $[2-2]_p$ will then happen at all plateaus except
the first two and the last two. There will also be second order
transitions of the type $[2-1]_p$ to $[2-1]_g$, between
$[2S-(j\!-\!2S)]_p$ and \mbox{$[2S-(j\!-\!2S)]_g$} for
\mbox{$(2S\!+\!1) \le j \le 4S$}. Thus we expect most features of the
phase diagram in Fig. \ref{phasediagram} to be generic for
all spins.

\section{Summary and conclusion}

We have generalized a class of existing spin-$\frac{1}{2}$ models in
various dimensions, which exhibit magnetization plateaus, to arbitrary
values of spin. By using simple techniques we are able to analytically
solve the spin-$S$ model in a substantial regime of the parameter
space. Our approach also throws light upon the phases beyond the exact
regime. In particular, we are able to predict that there will be a new
type of plateau phase. Such phases cannot exist for the
spin-$\frac{1}{2}$ ladders studied in HMT. We have tested our results
numerically for the particular case of $S=1$ two-leg ladder.

Though our general model is defined on arbitrary bipartite lattices,
each site has a sub-structure which includes $n$ spins. Therefore,
constructing physically feasible models, where bonds of equal strength
have equal length, is a non-trivial task. As far as we know, the
one-dimensional linked tetrahedra introduced by Gelfand\cite{Gelfand}
and Sutherland's generalization of the same to $3$-$d$ are the only
possibilities.

\section{Acknowledgments}
We thank Diptiman Sen and R.~Shankar for very useful discussions.
V.R.C. thanks Council for Scientific and Industrial Research, India for
a Junior Research Fellowship. N.S. acknowledges Department of Science
and Technology, India for financial support through Grant
No. SP/S2/M-11/2000.



\begin{thebibliography}{199}

\bibitem{CabraGrynbergHonecker} D.C.~ Cabra, M.D.~Grynberg, A.~
Honecker, P.~Pujol, Condensed Matter Theories (Vol {\bf 16}),
Ed. S.~Hernandez and J.W.~Clark (New York: Nova Science), 17 (2001);
Preprint cond-mat/0010376, and references therein.

\bibitem{SchulenburgHoneckerSchnack} J.~Schulenburg, A.~Honecker,
  J.~Schnack, J.~Richter, H.-J.~Schmidt, Phys. Rev. Lett. {\bf 88},
  167207 (2002).

\bibitem{Gelfand} M.P.~Gelfand, Phys. Rev. B {\bf 43}, 8644 (1991).

\bibitem{BoseGayen} I.~Bose, S.~Gayen, Phys. Rev. B {\bf 48}, 10653 (1993).

\bibitem{MullerSinghKnetter} E.~M\"{u}ller-Hartmann,
  R.R.P.~Singh, C.~Knetter, G.S.~Uhrig, Phys. Rev. Lett. {\bf 84},
  1808 (2000).

\bibitem{HoneckerMilaTroyer} A.~Honecker, F.~Mila, M.~Troyer, Eur. Phys. J. B
  {\bf 15}, 227 (2000).

\bibitem{Sutherland} B.~Sutherland, Phys. Rev. B {\bf 62}, 11499
  (2000).

\bibitem{Mila} F.~Mila, Eur. Phys. J. B {\bf 6}, 201 (1998)
 
\bibitem{ShastrySutherland} B.S.~Shastry, B.~Sutherland, Physica B
  {\bf 108B}, 1069 (1981).

\bibitem{Bethe} H.A.~Bethe, Z. Phys. {\bf 71}, 205 (1931).

\bibitem{Haldane} F.D.M.~Haldane, Phys. Lett. {\bf 93A}, 464 (1983);
  Phys. Rev. Lett. {\bf 50}, 1153 (1983).

\bibitem{ParkinsonBonner} J.B.~Parkinson, J.C.~Bonner, Phys. Rev. B
  {\bf 32}, 4703 (1985).

\bibitem{LouQinSu} Jizhong Lou, Shaojin Qin, Zhaobin Su, Phys. Rev. B
  {\bf 62}, 13832 (2000).


\end{thebibliography}
\end{document}